\documentclass[a4paper]{article}

\usepackage{todonotes}
\newcounter{annecounter}

\newcounter{silviacounter}

\newcounter{chriscounter}

\newcounter{saracounter}

\usepackage{INTERSPEECH2021}

\title{Low-dimensional representation of infant and adult vocalization acoustics}
\name{Silvia Pagliarini$^1$, Sara Schneider$^2$, Christopher T. Kello$^2$, Anne S. Warlaumont$^1$ %\anne{We may have co-authors to add who would also give some feedback on the paper and project, but this is still TBD.}
}
\address{
  $^1$Department of Communication, University of California, Los Angeles, USA
  $^2$Cognitive and Information Sciences, University of California, Merced, USA}
\email{pagliarini@ucla.edu, sschneider2@ucmerced.edu, ckello@ucmerced.edu, warlaumont@ucla.edu}

\begin{document}

\maketitle
\begin{abstract}
During the first years of life, infant vocalizations change considerably, as infants develop the vocalization skills that enable them to produce speech sounds. Characterizations based on specific acoustic features, protophone categories, or phonetic transcription are able to provide a representation of the sounds infants make at different ages and in different contexts but do not fully describe how sounds are perceived by listeners, can be inefficient to obtain at large scales, and are difficult to visualize in two dimensions without additional statistical processing. Machine-learning-based approaches provide the opportunity to complement these characterizations with purely data-driven representations of infant sounds. Here, we use spectral features extraction and unsupervised machine learning, specifically Uniform Manifold Approximation (UMAP), to obtain a novel 2-dimensional spatial representation of infant and caregiver vocalizations extracted from day-long home recordings. UMAP yields a continuous and well-distributed space conducive to certain analyses of infant vocal development. For instance, we found that the dispersion of infant vocalization acoustics within the 2-D space over a day increased from 3 to 9 months, and then decreased from 9 to 18 months. The method also permits analysis of similarity between infant and adult vocalizations, which also shows changes with infant age.
\end{abstract}

\section{Introduction}
Human infant vocal repertoires undergo dramatic, multifaceted, and linguistically significant changes over the first two years of postnatal life. At birth, infants produce simple precursors to speech sounds, predominantly quasivowels and other short, quiet sounds~\cite{oller2000emergence, buder2013acoustic}. Over subsequent months, the sounds they produce diversify and become more complex, to include a range of loudness, vocal qualities, pitches and pitch contours, durations, and primitive consonant productions. By about 7 months, they begin to produce canonical babbles (which contain speech-like syllables combining consonants and vowels); over the following months and years, additional changes in infant vocal repertoires and vocal motor control continue~\cite{oller2000emergence,nathani2006assessing,vihman1996phonological}. This learning provides an essential foundation for linguistic communication, and is believed to be supported both by intrinsically-motivated play and by contingent and often imitative interactions with adult caregivers~\cite{bloom1988quality,papouvsek1989forms,kuhl2004early,goldstein2008social,moulin2014self,long2020social,warlaumont2020infant}. Infant language acquisition in general is strongly influenced by auditory inputs from caregivers and other environmental sound sources (e.g.,~\cite{papouvsek1994melodies,kuhl2004early,gervain2008infant,fusaroli2019hearing}). An infant's auditory environment includes voices from caregivers and other individuals of differing ages and genders plus a wide range of other sound sources, such as animals, physical objects, and electronic devices. The presence of these many different types of sound sources, the high variety of sounds produced by each, and the wide range of sound types produced by infants themselves present a challenge for characterizing infant vocal productions and auditory inputs.

Most characterizations of infant vocalizations are based on acoustic analyses of specific acoustic features (such as fundamental frequency and formant frequencies (e.g.,~\cite{kent1982acoustic,ritwika2020exploratory}), human listener categorizations into protophone categories (e.g. quasivowel, canonical babble, etc.) \cite{oller2000emergence,nathani2006assessing,buder2013acoustic}, or phonetic transcription \cite{vihman1996phonological}.  A limitation of these methods is that they don't provide a fully comprehensive characterization of a sound. For example, two sounds may be similar in their pitch characteristics but differ in terms of the phonetic features they contain. Or the sounds may belong to the same protophone category but still acoustically differ considerably and be perceived differently by listeners. Data-driven analyses of raw acoustic information could provide a complementary approach.

Sainburg et al.~\cite{sainburg2020finding} showed the power of using Uniform Manifold Aproximation~(UMAP)~\cite{mcinnes2018umap}, a machine-learning method for reducing a high-dimensional acoustic dataset into a two-dimensional space, to represent human speech sounds as well as songbird syllables. For human speech, different phonetic categories tended to project to distinct regions of latent space. For birdsong, the repertoires of different songbird species could be represented in the same overall space, enabling comparisons across species. %\anne{It might be helpful to add a couple sentences to the intro. explaining what UMAP and tSNE are and how they work and providing one or two references for each. We're out of time and space now at this point, but we should consider putting that in Silvia's ICIS talk and also in the longer manuscript we plan to write extending on this short paper.} 

Here we apply spectral features extraction and UMAP to transform raw acoustic data from daylong home audio recordings of 3- to 18-month-old infants. This generates a two-dimensional spatial representation of infant and caregiver vocalizations. We then use this space to quantify, for each daylong recording (1)~the similarity between infant and caregiver vocalizations and (2)~the diversity (i.e., the amount of variation) across infant vocal productions on the day of recording. Finally, we assess how these measures vary with infant age. %\anne{We can save space by skipping the overview of what the different sections contain. (Feel free to delete this comment once you've read it, if you agree.)}
% Section~\ref{methods} describes the data pre-processing, how to build the vocalizations space, and the statistical analysis we performed to explore it. Section~\ref{results} shows the vocalizations space and the results obtained from a dataset of long-day recordings of families obtained at $3$, $6$, $9$, and $18$ months. Section~\ref{discussion} summarizes our findings, and discuss their limitations and perspectives.

\section[methods]{Methods\footnote{More details are available at https://github.com/spagliarini/Infant-vocalization-space-Interspeech2022}}
\subsection{Dataset and preprocessing}
%\silvia{add name of the dataset and reference? if we keep validation in results, add subsec with complete and homebank} \anne{I added those details, thanks for flagging the need. Also, for future reference I prefer to call the homebank-child-voc-cleaning data ``cleaned'' or ``validated'' because ``homebank'' just refers to the HomeBank data sharing project (homebank.talkbank.org) which has a lot of other components. So if you say ``homebank data'' to me I get confused. I realize now in retrospect I could have named the repository something better!}
Our infant and adult vocalization data comes from long-form (10+ daytime hours), child-centered home recordings. $52$ infants were recorded longitudinally at $3$, $6$, $9$ and $18$ months using the LENA system~\cite{gilkerson2008lena}. A few infants did not complete all four recordings; we focused on the subset of $42$ infants who did have complete data. More details about the data collection methods can be found in \cite{ritwika2020exploratory}. A subset of the recordings are available in the Warlaumont corpus \cite{warlaumont2016warlaumont} within HomeBank \cite{vandam2016homebank}. We extracted short audio clips of duration $0.6s$ to $12s$ for sections of the recording labeled by LENA's built-in algorithm~\cite{xu2008signal} as one of the following four types: class \textit{CHNNSP} contains cry/laugh (non-speech-related) vocalization produced by the infant wearing the recorder, class \textit{CHNSP} (child speech-related) contains speech or protophone vocalizations produced by the infant wearing the recorder, class \textit{FAN} (female adult near) contains adult female vocalization, and class \textit{MAN} (male adult near) contains adult male vocalizations. Although LENA also tags other sound sources (noise, television, other adults, other children, overlap), for the present study we included only the infant 
%wearing the recorder 
and their adult caregivers. Table \ref{tab:datasets} gives the number of instances of each class across the 42 participants. %\anne{The table is for the 42 infants, not the 52, correct?}. 

\begin{table}[th]
\footnotesize
\centering
\caption{Total utterances for each LENA sound category}
  \label{tab:datasets}
\begin{tabular}{lllll}
\hline
Age(months) & CHNNSP & CHNSP & FAN & MAN \\ \hline
3  & 32631 & 39110 & 84999 & 35170 \\
6  & 29928 & 43932 & 70491 & 29674 \\ 
9  & 32007 & 45246 & 69802 & 29629 \\ 
18 & 32852 & 60024 & 69351 & 31949 \\ \hline
\end{tabular}
%\anne{This is for the 42-participant dataset? (Also, the top row of the table could probably just be incorporated into the table caption.)} \silvia{Yes, it is for the 42. Do you mean that you would write it as text? or just remove the top line?}
\end{table}

\subsection{Low-dimensional vocalization space} \label{methods_UMAP}

%We combined spectral feature extraction and unsupervised machine learning to represent the vocalizations in a low-dimensional space. In such a space, sound clips with similar acoustic features are positioned near each other.

\subsubsection{Spectral feature extraction}
We used openSMILE~\cite{eyben2010opensmile} to extract $13$ Mel-frequency cepstral coefficients (MFCCs) from $26$ Mel-frequency bands (with $25 ms$ frame size and $10 ms$ frame rate). Each MFCC's first and second derivatives (i.e., velocity and acceleration) were also computed, generating an additional 26 features. This gave us, for each child or adult vocalization clip, a $39 \times T$ matrix where $T$ was the number of timebins in the clip. We then summed across each matrix's timebins, yielding a 39-dimensional vector for each sound clip.

\subsubsection{Dimensionality reduction}
From collections of these feature vectors 
%\anne{We may want to put in results which collections are relevant and were included in the UMAP analysis for that particular analysis, if it's not already there.} 
we obtained a 2D representation of the vocalization space using Uniform Manifold Approximation and Projection for Dimension Reduction (UMAP) \cite{mcinnes2018umap}. Similar to t-distributed stochastic neighbor embedding (t-SNE~\cite{van2008visualizing}), UMAP is a dimension reduction technique where the result axes do not represent single, meaningful discriminant features. In particular, UMAP (1) can be used to perform non-linear dimensionality reduction, (2) has a higher computational efficiency than t-SNE, and (3) can be non-local (i.e., it can take into account distances between points that are located far apart from each other). The tuning of UMAP hyperparameters enables more or less local representation of the data. A smaller neighborhood size $n_{neigh}$ means a more locally-focused representation; larger values will push UMAP to look at larger neighborhoods of each point when estimating the manifold structure of the data. The minimum distance $min_{dist}$ parameter determines how far points are allowed to be from each other in the low dimensional representation; this also influences the balance of emphasis on local detail vs. broad structure. We used the UMAP Python package\footnote{https://github.com/lmcinnes/umap}~\cite{mcinnes2018umap}.%, which depends upon scikit-learn and its dependencies. 
We chose $n_{neigh} = 15$ and $min_{dist}=0.1$.
%\anne{Could note that these are the defaults (although Silvia did do some informal explorations) and that they are also what was used for Silvia's canary song thesis. Since we're out of space and time, let's plan to do this for the journal submission. Actually, it would be nice also to see if our results change or are robust to different parameter values.}

\subsubsection{Statistical analyses}
For each daylong recording, we measured (1) the distance from the centroid of the infant vocalizations to the centroid of the FAN vocalizations (we focused our statistical analyses in this initial study on CHNSP-to-FAN distance because some recordings had extremely few MAN vocalizations), (2) the average distance from the infant vocalizations to the infant's centroid and (3) the Shannon entropy of \textit{CHNSP} vocalizations.

%For each infant at a particular age, the centroid coordinates $C_{x,y}^L$, where $L=\{CHNNSP, CHNSP, FAN, MAN\}$ in the 2D vocalizations space obtained via UMAP %were calculated as follow:
%\begin{equation*}
%    C_{x,y}^L = \bigg( \frac{\Sigma_{i=1,..,N^L} x_i^L}{N^L}, \frac{\Sigma_{i=1,..,N^L} y_i^L}{N^L} \biggl),
%\end{equation*}
%where $N^L$ is the number of vocalizations contained in class $L$, $x_i^L$ represents the value along the first axis of the UMAP representation for vocalization $i$ of class $L$, and $y_i^L$ represents the value along the second axis.

We then performed (in R, using lme4 \cite{bates2015fitting} and {lmerTest} \cite{kuznetsova2017lmertest}) linear mixed effects regressions with infant ID as a random effect and the number of infant vocalizations in the daylong recording as a covariate to test the hypothesis that infant age influences (1) the distance between the centroid of the infant vocalizations and the centroid of the adult female vocalizations, (2) the average distance from individual infant vocalization locations to the infant vocalization centroid, and (3) the Shannon entropy of the infant vocalization locations. We included both linear and quadratic terms for age in each regression.

\section{Results}
\label{results}
\subsection{UMAP-based low-dimensional vocalizations space}
\begin{figure*}
    \centering
    \includegraphics[scale=.25, trim={3cm 17cm 0cm 0cm}]{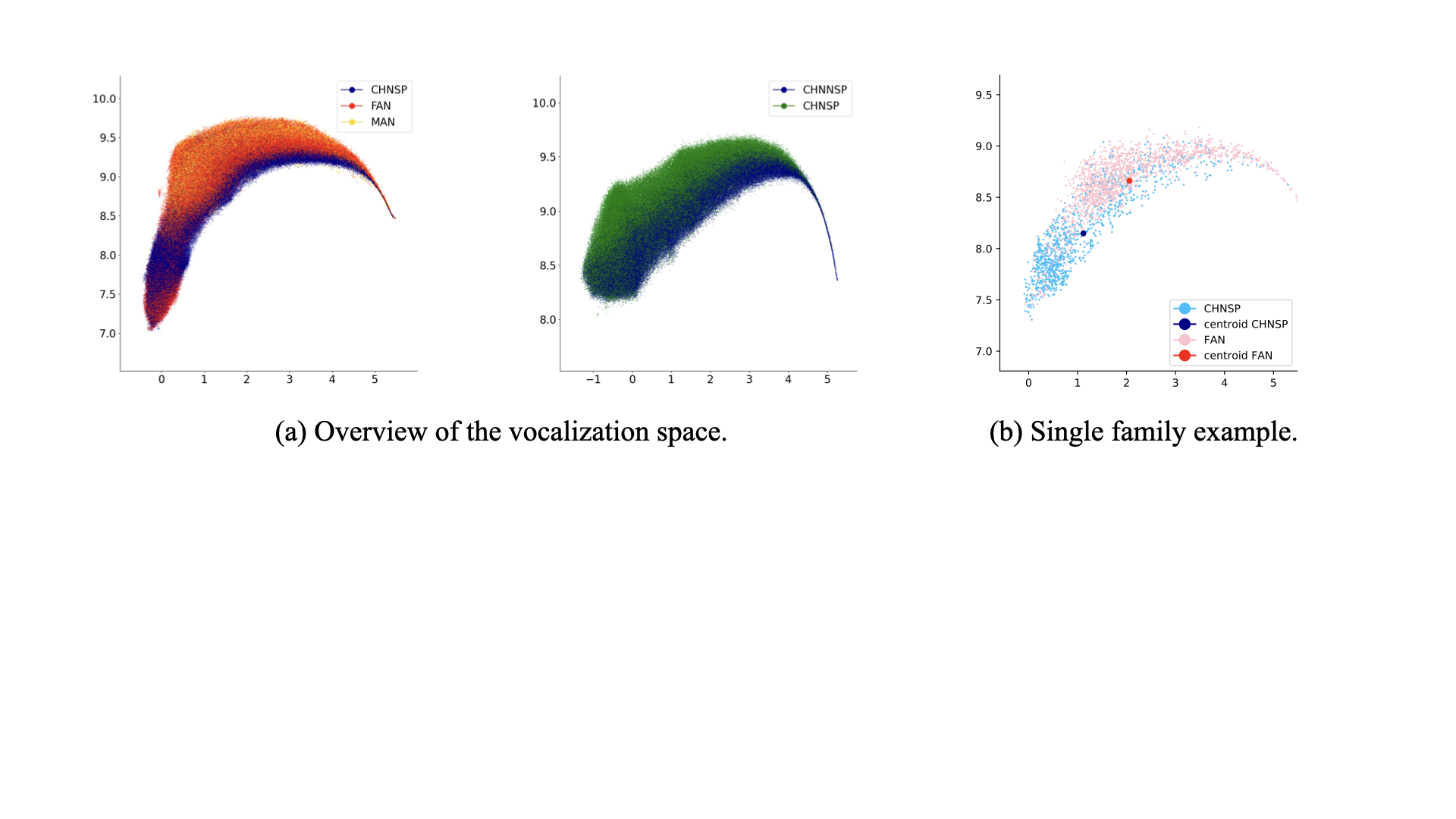}
    \caption{\textbf{UMAP-based vocalization space.} Representation of all vocalization clips in a space generated by applying UMAP to MFCC features summed across the clip. Each point represents a vocal utterance as 39-dimensional vector ($13$ MFCCs, $13$ MFCC velocities, and $13$ MFCC accelerations). (a) UMAP space obtained from non cry/laugh infant, adult male and female vocalization clips. Each color represents a class: \textit{CHNSP} are infant speech and pre-speech non cry/laugh vocalizations (blue dots), \textit{MAN} and \textit{FAN} are respectively female (red dots) and male (yellow dots) adult vocalizations, \textit{CHNNSP} are infant cry/laugh or vegetative%(non-speech-related) 
    . The left panel shows the main space analysed in this paper, constructed based on \textit{CHNSP}, \textit{FAN}, and \textit{MAN} clips. The right panel shows a space constructed based on \textit{CHNNSP} and \textit{CHNSP}. (b) Example of a single 18-months old baby recording (extracted from the whole UMAP space). The light blue dots represent the infant vocalizations (\textit{CHNSP}), the pink dots represent the adult female vocalizations (\textit{FAN}). The centroid for class \textit{CHNSP} is shown by the dark blue dot and the centroid for class \textit{FAN} is the red dot.} %\anne{For the left plot in (a), are the CHNNSP included in the space calculation, just left out from the plot for clarity? Or are they not included in the space calculation?} \silvia{they are not included in this space}
    \label{fig:results_UMAP}
\end{figure*}
%\silvia{I just realized that we use the word vocalization all over this section, and possibly before/after, while referring to the MFCC sum of each vocalization. Can this be measleading? We can change and use another word, or simply say in the methods that we use the word vocalization to refer to the processed vocalization (could be added at the end of section 2.2.1.}
%\anne{I added `` summed across the clip'' to the figure caption...we can check the rest of the paper and add similar phrases to make sure it's clear and precise, as needed.}

Vocalizations clustered together according to the sound type classes, but with a large amount of overlap. This can be observed in the vocalization space obtained from the collection of infant speech-related (\textit{CHNSP}), adult male (\textit{MAN}) and adult female vocalizations (\textit{FAN}) (left panel of Fig.~\ref{fig:results_UMAP}a; our analyses below focus on this space) and in a vocalizations space obtained from the collection of both types of infant sounds (\textit{CHNSP} and \textit{CHNNSP}) (right panel of Fig.~\ref{fig:results_UMAP}a).
Similarly, in a space constructed from a single daylong recording, vocalizations cluster together when comparing the \textit{CHNSP} and \textit{FAN} vocalizations (Fig.~\ref{fig:results_UMAP}b). Figure~\ref{fig:results_UMAP}b also illustrates the centroids of each class.

\subsection{Similarity between infant and adult female vocalizations}

\begin{figure*}
    \centering
    \includegraphics[scale=.28, trim={0cm 0cm 0cm 0cm}]{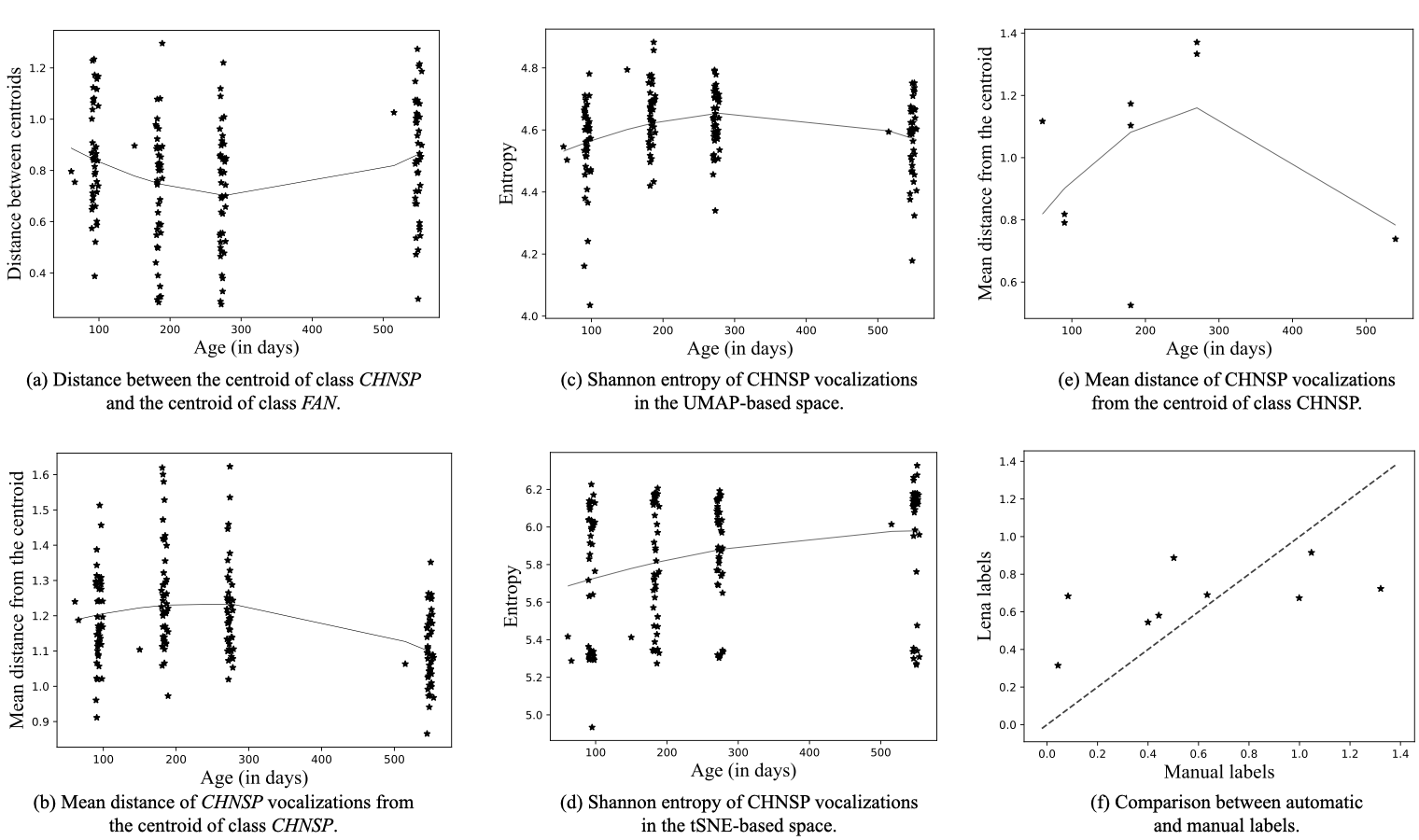}
    \caption{\textbf{Age-related changes in UMAP-based measures, comparisons to tSNE, and results for human-validated infant utterances.} Each star represents data from a particular infant's recording at a given infant age. (a) Similarity between infant speech-related (\textit{CHNSP}) and adult female (\textit{FAN}) vocalizations plotted as a function of age. (b) Variability of \textit{CHNSP} vocalizations in the UMAP-based vocalizations space, measured as distances to the \textit{CHNSP} centroid, as a function of age. (c) Shannon entropy of CHNSP vocalizations in the UMAP-based vocalizations space versus age. (d) Shannon entropy of CHNSP vocalizations in a tSNE-based vocalizations space versus age. (e) Variability of \textit{CHNSP} vocalizations in the UMAP-based vocalizations space for a subset of the recordings with child vocalizations verified as such by human listeners. (f) Scatter plot showing the mean distance from centroid measure (for \textit{CHNSP}) when child vocalizations were identified automatically by LENA ($y$-axis) versus when those \textit{CHNSP} clips were manually verified by human listeners as being infant sounds and not mis-classifications by the labeling algorithm ($x$-axis). The black lines in panels (a-e) represent data fitting obtained using polynomial regression on the represented data. The dashed line in (f) represents hypothetical 1:1 correspondence between the two labeling methods.} %\silvia{change and add the one with smooth line}}
    \label{fig:results_statistical}
\end{figure*}

We assessed the similarity between an infant's vocalizations on a given day and the corresponding adult female vocalizations that infant was exposed to %on the same day 
by calculating the distance from the centroid of the infant vocalizations to the centroid of the adult female vocalizations. There was a statistically significant positive quadratic effect of age on similarity between infant and adult female ($p<.001$), with  a decrease in centroid distance from 3 to 9 months followed by an increase at 18 months (Figure~\ref{fig:results_statistical}a). This suggests that infant and adult female vocalizations become increasingly similar from $3$ to $9$ months of age, and then diverge again by the time infants get to be $18$ months old.

\subsection{Infant vocalization variability}

For each infant recording, we assessed infant vocalization variability by calculating the average distance of the infant's \textit{CHNSP} vocalization locations from the class centroid in the UMAP-generated space. We then asked if this variability changed across infant age (Fig.~\ref{fig:results_statistical}b). Statistical analyses detected both a linear ($p < .001$) and quadratic ($p < .001$) relationship between age and infant vocalization variability. Both effects were negative, indicating that the diversity of sounds infants produced during a day rose to a peak at $9$ months and then decreased from then on, with the lowest variability observed at $18$ months. Shannon entropy analysis (Fig.~\ref{fig:results_statistical}c) of the infant non cry/laugh vocalizations replicated the inverted-U-shape observed in the centroid analysis (Fig.~\ref{fig:results_statistical}b), with a statistically significant negative quadratic term ($p < .001$).

\subsection{tSNE-based vocalizations space}
\begin{figure}[h]
    \centering
    \includegraphics[scale=.32]{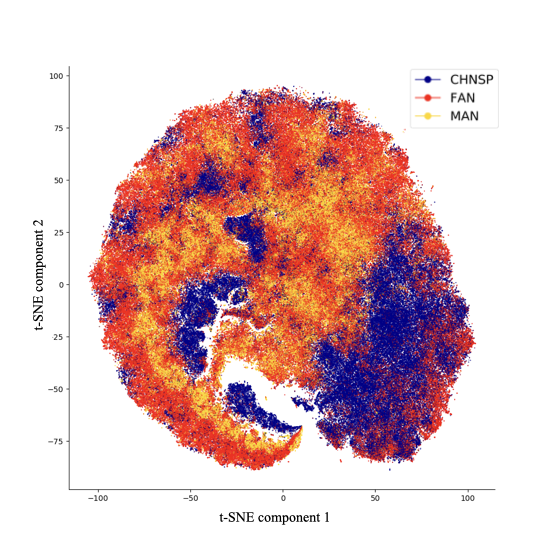}
    \caption{\textbf{tSNE-based vocalizations space.} Representation of \textit{CHNSP} (blue), \textit{FAN} (red), and \textit{MAN} (yellow) vocalization clips in a space generated by applying tSNE to MFCC features summed across each clip. Each point represents a vocal utterance as 39-dimensional vector ($13$ MFCCs, $13$ velocities, and $13$ acceleration). Vocalizations space obtained from non cry/laugh infant, adult male and female vocalization clips. Each color represents a class: \textit{CHNSP} are infant speech and pre-speech non cry/laugh vocalizations (blue dots), while \textit{MAN} and \textit{FAN} are respectively female (red dots) and male (yellow dots) adult vocalizations.}
    \label{fig:results_tSNE_space}
\end{figure}

As mentioned in Section~\ref{methods_UMAP}, tSNE is another method that can be used to perform non-linear dimensionality reduction. The space obtained using tSNE in place of UMAP for the \textit{CHNSP}, \textit{FAN}, and \textit{MAN} classes is shown in Figure~\ref{fig:results_tSNE_space}. This figure can be compared with the representation of the same data obtained with UMAP (statistically significant quadratic term: $p<.001$), shown in the left panel of Figure~\ref{fig:results_UMAP}a. Interestingly, Shannon entropy analysis (Fig.~\ref{fig:results_statistical}d) of the \textit{CHNSP} vocalizations in the tSNE-generated space ($p=0.03$) do not appear to have a clear U-shape with respect to age, as was observed in the UMAP case (Figure~\ref{fig:results_statistical}(b-d)). (Given the high discontinuity of the tSNE-based space, we do not compare centroid-based measurements.)

\subsection{Validation of the dataset}
\label{section:validation}
For a subset of our data ($9$ recordings in total), we worked with a team of ten human listeners to re-label the automatically identified \textit{CHNSP} clips. Listeners assigned a prominence value $P \in \{1,2,3,4,5\}$ to indicate if they believed the infant was indeed vocalizing in the clip without any other sounds audible ($P=1$), if the infant was present but so were other sounds and if so, what was the infant sound prominence relative to those other sounds background noise and how much ($1<P<5$), or if the infant wearing the recorder did not actually vocalize at all during the clip ($P=5$). For each clip, we computed the modal value across listeners and defined a threshold for inclusion in the strictest way possible: we validated a clip as infant prominent if its modal value was equal to $1$. Otherwise, we considered the clip to be noise and excluded it from the study. We then computed the mean distance of infant vocalizations (\textit{CHNSP}) from the class centroid (Figure~\ref{fig:results_statistical}e), and observed a comparable trend with what we obtained from the whole dataset (Figure~\ref{fig:results_statistical}b). We also observed that the range of this variability measure was larger for the human-validated labels than for the fully automated LENA labels (Fig.~\ref{fig:results_statistical}f). The correlation between the two sets of labels is equal to $0.47$.

\section{Discussion}
\label{discussion}

We proposed a method based on spectral features extraction (MFCC's + their first and second derivatives, summed across timebins for an utterance) and UMAP~\cite{mcinnes2018umap}, an unsupervised machine learning dimensionality reduction method, to represent infants and adult vocalizations in a two-dimensional space. For our dataset, this resulted in a relatively smooth distribution of the data across space (in contrast with tSNE). Statistical analyses revealed significant non-monotonic patterns of change in age. Specifically, (1) the range of sound patterns infants produced (as quantified by distance-to-class centroid and Shannon entropy) increased from 3 to 9 months and then decreased by the time infants were 18-months-old. We also found that similarity between infant and adult vocalizations increased from 3 to 9 months then decreased from 9 to 18 months.

We hypothesize that the increase from 3 to 9 months might relate to the so-called ``expansion stage'' of infant protophone development as well as to the onset of canonical babbling \cite{oller2000emergence}. This might be tested by establishing if there is a relationship between locations in the UMAP-constructed space and protophone categorizations of the infant sounds. We also note that previous work~\cite{kent1982acoustic} observed informally (without tests for statistical significance) that as infant age increased from 3 to 9 months, so did the range of infant vocalization durations and the formant frequency ranges. Our findings are consistent with those trends for increasing acoustic variation from 3 to 9 months, but also show that by 18 months, the amount of variation shows a decrease from its 9-month level. It would be informative (1)~to compare the relationship between adult female and infant non cry/laugh vocalizations with the relationship between adult female and infant cry/laugh vocalizations, and (2)~to connect the UMAP representations to more transparent and commonly used acoustic features, such as duration, formant frequencies, and pitch features.

A limitation is that LENA-generated sound source labels are sometimes incorrect, and even when they are correct, there are sometimes significant other sounds present~\cite{rasanen2021alice}. %This is due to the complexity and the noisiness of the data, and to the fact that LENA's core technology is aging with respect to current advances in automated speech processing \cite{rasanen2021alice}. %\silvia{Maybe we can/need add more refs}
To overcome this problem, we are working to obtain a cleaner dataset based on human listener judgments. Preliminary results (Section~\ref{section:validation}) suggest that the overall qualitative patterns may not change drastically, but additional work and data are needed.
%\silvia{if we keep the section about the validation, mention it here. Finish/modify this part depending on what we keep in the results}.

In the future, variations on this approach could be explored, including alternative pre-processing and dimensionality reduction methods. For instance, it could be useful to pre-process the audio using end-to-end neural network approaches, perhaps with a training goal of optimizing infant age predictions. This may allow UMAP to be performed on more sophisticated and more functionally-and practically-relevant acoustic features.
%It would also be worth exploring alternative ways of characterizing similarity and dispersion within the derived spaces (e.g. size of the clusters, correlation analysis, covariance of vocalizations over time). 
This general approach may be useful for characterizing the interactions between infants and caregivers, for studying individual and clinical differences in vocal productions, and for providing additional means for comparison of human infant behavior to that of computational models of vocal learning~\cite{pagliarini2020vocal}.

\section{Acknowledgements}
This work was supported by the National Science Foundation (grants 1529127, 1539129/1827744, and 1633722) and by the James S. McDonnell Foundation (doi: 10.37717/220020507). We thank Dr. Gina Pretzer, Sara Mendoza, Jimel Mutrie, Curt Chang, and all the research assistants who assisted with data collection, audio vetting, and child vocalization re-labeling.

\bibliographystyle{IEEEtran}

\newpage
\bibliography{interspeech2021_bib}

% \begin{thebibliography}{9}
% \bibitem[1]{Davis80-COP}
%   S.\ B.\ Davis and P.\ Mermelstein,
%   ``Comparison of parametric representation for monosyllabic word recognition in continuously spoken sentences,''
%   \textit{IEEE Transactions on Acoustics, Speech and Signal Processing}, vol.~28, no.~4, pp.~357--366, 1980.
% \bibitem[2]{Rabiner89-ATO}
%   L.\ R.\ Rabiner,
%   ``A tutorial on hidden Markov models and selected applications in speech recognition,''
%   \textit{Proceedings of the IEEE}, vol.~77, no.~2, pp.~257-286, 1989.
% \bibitem[3]{Hastie09-TEO}
%   T.\ Hastie, R.\ Tibshirani, and J.\ Friedman,
%   \textit{The Elements of Statistical Learning -- Data Mining, Inference, and Prediction}.
%   New York: Springer, 2009.
% \bibitem[4]{YourName17-XXX}
%   F.\ Lastname1, F.\ Lastname2, and F.\ Lastname3,
%   ``Title of your INTERSPEECH 2021 publication,''
%   in \textit{Interspeech 2021 -- 20\textsuperscript{th} Annual Conference of the International Speech Communication Association, September 15-19, Graz, Austria, Proceedings, Proceedings}, 2020, pp.~100--104.
% \end{thebibliography}

\end{document}